# Nova - Un nuage arc-en-ciel au-dessus des Alpes

**Nicolas Gibelin**
UMS 3758 GRICAD – Bâtiment IMAG – 700 avenue centrale – 38400 Saint Martin d'Hères

**Rémi Cailletaud**
UMS 832 OSUG – 122 rue de la piscine – 38 400 Saint Martin d'Hères

**Gabriel Moreau**
UMR 5519 LEGI – 1209-1211 rue de la piscine – Domaine Universitaire – 38400 Saint Martin d'Hères

**Jean-François Scariot**
INRIA Grenoble Rhône-Alpes – 655 Avenue de l'Europe - CS 90051 – 38334 Montbonnot Cedex

**Gabrielle Feltin**
UMS 3758 GRICAD – Bâtiment IMAG – 700 avenue centrale – 38400 Saint Martin d'Hères

**Anthony Defize**
UMS 3758 GRICAD – Bâtiment IMAG – 700 avenue centrale – 38400 Saint Martin d'Hères

## Résumé

*Un service d'infrastructure à la demande (IaaS) basé sur la suite logicielle OpenStack [8] a été déployé sur le campus universitaire de Grenoble durant l'année 2018 puis mis à jour en 2019.*

*Nous présentons les méthodes que nous avons choisies pour déployer et gérer l'infrastructure : Racadm et Preseed pour l'installation du système de base, puis Kolla [9] pour le déploiement d'OpenStack. Cette dernière solution, basée sur des conteneurs pour chaque service, permet une configuration centralisée et historisée (GitLab) des contrôleurs et des nœuds de calcul. La solution s'impose comme la solution de référence pour un déploiement reproductible d'OpenStack. Notre nuage a ainsi été facilement étendu avec de nouveaux nœuds. Le changement de version de l'OS de base a aussi pu être testé avec succès malgré quelques petits couacs… La sécurité étant un élément clef du bon fonctionnement de ce type de service mutualisé, nous avons fait en sorte que chaque projet ainsi que l'ensemble de ses données soit parfaitement isolé. Tous les flux réseaux sont chiffrés dans des VXLAN.*

*Cette plate-forme d'infonuagisme OpenStack est opérationnelle. À quoi tout cela sert-il ? Nos premiers utilisateurs font par exemple du Notebook Jupyter au travers de la mise à disposition de serveurs Jupyterhub (portail web), du Système Informatique Distribué d'Évaluation en Santé (projet SIDES), de l'intégration continue en liaison avec la plateforme GitLab, du test pour l'ordonnanceur de conteneurs Kubernetes ou encore du logiciel de calcul et de visualisation, etc, des usages très variés que les autres plateformes avaient des difficultés à proposer.*

## Mots-clefs

*OpenStack, Cloud Computing, Nuage, IaaS, Ansible, Kolla, Idrac, Racadam.*

# 1   Introduction

Les établissements du campus Grenoblois disposent de plusieurs plateformes de calcul haute performance, dont une grosse plateforme de virtualisation sous VMware. Ces plateformes ne sont néanmoins pas adaptées à tous les usages, soit par leur difficulté de prise en main (gestionnaires de jobs, environnements contraints, *etc.*), soit par le temps de déploiement d'un service.

De plus, les laboratoires de Santé et de SHS, dynamisés par la fusion de l'Université de Grenoble Alpes en 2016, ont un fort besoin de collecte et de traitement de données massives. Toutefois, ces structures manquent de ressources humaines et matérielles. Les machines de calculs de l'UMS GRICAD [4] (mésocentre régional Ciment) orientées HPC et/ou pour les séries de calculs séquentiels, ne sont actuellement pas adaptées à l'utilisation d'outils comme ElasticSearch, Neo4j, *etc*.

Plusieurs formations universitaires sont intéressées par une plateforme d'infrastructure à la demande. En Mathématiques Appliquées, Informatiques et Géosciences, les étudiants doivent avoir leur propre environnement et réaliser des TP sur machines virtuelles. Ces filières utilisent aussi la plateforme mutualisée Jupyter. Historiquement, celle-ci était hébergée sur un serveur physique. Un des objectifs du projet était de transférer celle-ci dans une infrastructure de nuage, afin d'équilibrer dynamiquement le service en fonction de la charge et des examens !

Suite à ce constat, le service d'infrastructure à la demande (IaaS) est né, porté par l'UMS GRICAD, dédiée aux infrastructures de calcul intensif et de données. Cette plate-forme d'infonuagisme basée sur la suite logicielle OpenStack [8] est maintenant opérationnelle et monte progressivement en charge.

# 2   Le projet

## 2.1   La réflexion

Pour ce projet de cloud, nous avions dès le départ fait le choix d'utiliser des logiciels open source. OpenStack est l'une des plus grosses communautés open source au monde [11]. La modularité et la richesse fonctionnelle de cette suite logicielle sont les principaux arguments qui ont conduit à son choix.

Une des principales difficultés est le déploiement d'une infrastructure logicielle. Comment procéder en activant la HA sur tous les services, exploiter et mettre à jour les logiciels ? Comment gérer l'insertion et la suppression de matériels dans l'infrastructure ? Nous devions en plus prendre en compte des contraintes de temps limité, réduire les interruptions de service, et avec un nombre restreint de personnels.

Dans un premier temps, nous avons exploré des solutions intégrées. Nous avons échangé avec le constructeur Dell pour bénéficier de son expérience dans le déploiement des plate-formes de cloud. Une solution intéressante était celle de l'éditeur Bright Computing permettant une automatisation complète de la gestion du cloud. Elle permet l'installation automatique des systèmes d'exploitation et la configuration des services OpenStack. Toutefois, le prix des licences étaient élevés au regard de l'enveloppe budgétaire pour le socle. Elle ne nous permettait pas d'acheter le matériel.

Nous voulions garder le contrôle de l'installation et de la configuration. Toutefois, OpenStack exécute des dizaines de services en interne pour effectuer ses propres tâches avec un système d'élection lors des choix afin d'assurer la robustesse. Il nous a semblé très difficile d'installer chaque service manuellement [5][6]. Il était nécessaire de disposer d'une procédure automatique pour déployer et s'éloigner autant que possible des problèmes liés aux innombrables fichiers de configuration.

Nous avons alors cherché des solutions d'installation par outils de déploiement open source. Ansible apparaît rapidement dans ces outils de gestion de cloud comme une référence. En approfondissant le sujet, nous avons découvert le projet Kolla [9]. Des tests locaux convaincants, puis un déploiement multi-serveurs d'un cloud OpenStack complet nous ont permis de valider et d'adopter les outils de déploiement du projet Kolla.

Il manquait encore une phase importante du déploiement (que Bright Computing prenait en charge) : l'installation automatique du système d'exploitation sur les serveurs physiques. Les outils / APIs de Dell sur la technologie de l'iDrac embarquée dans les serveurs nous ont permis de résoudre ce problème.

La solution logicielle choisie et testée, nous avons alors conçu la base de notre architecture matérielle.

## 2.2 L'architecture choisie

Les principaux établissements Grenoblois se sont investis dans le projet : L'Université Grenoble Alpes, Grenoble INP, le CNRS et INRIA Rhône-Alpes. La participation financière initiale de chacun et du projet SIDES a permis de démarrer avec un investissement de 140 000 euros, incluant la partie réseau. L'UMS GRICAD, ayant en charge la mise à disposition d'infrastructures mutualisées orientées vers la recherche, a piloté ce projet. Un comité technique a été monté suivant un principe bien établi sur le site : un appel à toute la communauté à participer. Ainsi une dizaine de personnes d'horizons divers ont contribué à ce projet, lui permettant de passer de concept à réalisation collective concrète.

Pour la gestion des ressources physiques, la plateforme devait pouvoir passer facilement à l'échelle et s'appuyer sur du matériel performant et maîtrisé par les personnes du site. L'architecture OpenStack se base sur la séparation d'un plan de contrôle et de travail. Il est conseillé de déployer au moins trois machines pour le plan de contrôle [12]. Les trois contrôleurs reposent sur des serveurs DELL R640. Chaque contrôleur est connecté à un commutateur réseau différent par quatre ports 10 Gb/s (ACI) dédiés à des trafics indépendants (tunnel, stockage, API, trafic projets).

- Les nœuds de calcul sont hétérogènes. L'objectif de la plateforme est d'ajouter des nœuds à mesure des besoins. Nous avons commencé avec deux nœuds DELL R640, puis les anciens serveurs Jupyter (R730) ont rejoint la plateforme ainsi qu'un R640 (avec plus de mémoire) acheté par le projet SIDES. À ce jour, la plateforme de calcul compte donc 144 cœurs et 1,6 To de RAM.

- Un stockage Ceph [10] (Luminous) a été déployé en mode bloc afin d'y héberger les données « pérennes » des machines virtuelles. Ce cluster est composé de 3 contrôleurs R640 et de 3 nœuds de stockage R740xd portant sa capacité utile à 48 To. Les contrôleurs sont inter-connectés en 2x10 Gb/s et les nœuds de stockage en 4x10 Gb/s.

- Un second stockage est disponible sur une baie NetApp SolidFire (full SSD) de 8,6 To.

- Les données et résultats à longue durée de vie sont stockés sur la plateforme de stockage de site SUMMER [3] (baies NetApp).

- L'interconnexion de tous ces éléments nécessite un réseau très performant. Nous profitons dans le data-centre de la solution SDN (Software Design Network) ACI de CISCO déployée à l'UGA, et opérée par Spring [1]. Les différents éléments de la solution sont actuellement répartis sur quatre Leafs 48 ports 10 Gb/s inter-connectées entre elles à 40 Gb/s afin d'assurer une redondance et pallier toute défaillance.

La sécurité étant un élément clef du bon fonctionnement de ce type de service, nous avons fait en sorte que chaque projet ainsi que l'ensemble de ses données soit parfaitement isolé.

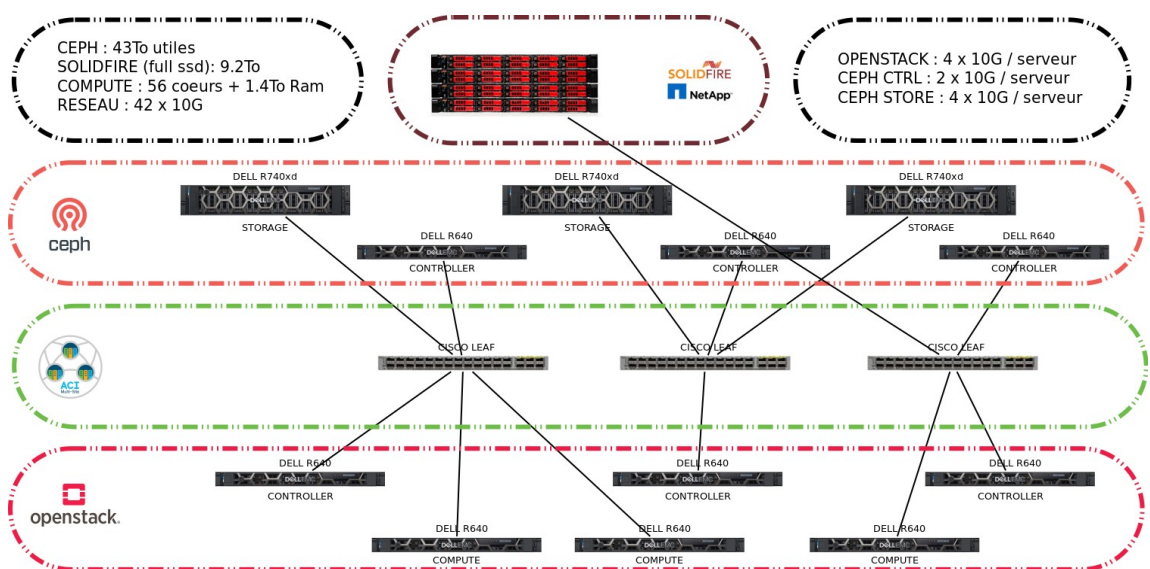

*Figure 1 - Architecture Nova*

## 2.3 Déploiement des bare-metals

Afin d'automatiser le plus possible de déploiement de la plateforme, nous avons décidé d'utiliser les capacités des iDrac de Dell pour l'installation des OS, et la technologie de « Preseed » pour l'installation automatique des distributions Ubuntu.

### 2.3.1 Configuration du Preseed

Preseed permet d'automatiser la réponse aux questions de l'installeur, sans avoir à les entrer manuellement. Il est alors possible de configurer l'installation complète, de la configuration réseau au partitionnement en passant par la définition de la locale du clavier.

Il est aussi possible d'exécuter un script Shell en toute dernière phase d'installation.

```
d-i preseed/late_command string \
   mkdir -p /target/root/preseed/; \
   cp -R /cdrom/preseed/* /target/root/preseed/; \
   in-target /bin/sh /root/preseed/run.sh
```

Attention : lors d'une installation sur une machine préalablement installée, l'installeur monte automatiquement les partitions présentes, ce qui peut poser problème. L'ajout des lignes suivantes, permet de le contourner :

```
d-i partman/early_command string \
USBDEV=$(list-devices usb-partition | sed "s/\(.*\)./\1/");\
BOOTDEV=$(list-devices disk | grep -v "$USBDEV" | head -1);\
debconf-set partman-auto/disk $BOOTDEV;\
debconf-set grub-installer/bootdev $BOOTDEV; \
umount /media; \
return 0;
```

Enfin, il faut déposer dans le répertoire `/isolinux` des fichiers `isolinux.cfg` :

```
# D-I config version 2.0
# search path for the c32 support libraries (libcom32, libutil
etc.)
path
include menu.cfg
default vesamenu.c32
prompt 0
timeout 1
ui gfxboot bootlogo
```

et `txt.cfg` :

```
default autoinstall
label autoinstall
  menu label ^Install Ubuntu Server for Openstack
  kernel /install/vmlinuz

  append auto=true file=/cdrom/preseed/openstack.seed vga=788
debian-installer/locale=en_US.UTF-8 keyboard-
configuration/layoutcode=fr console-setup/ask_detect=false
debian-installer/quiet=true debian-installer/framebuffer=false
initrd=/install/initrd.gz nosplash ---
```

Lors d'une installation manuelle, le fichier de configuration de l'installateur peut être récupéré :

```
debconf-get-selections –installer
```

### 2.3.2 Génération des images

Pour chaque serveur, nous pouvons créer une image d'installation, avec par exemple l'écriture « en dur » des configurations réseau de chaque hôte, pour éviter les problèmes de bail DHCP qui ne se renouvelle pas. On utilise pour cela un remplacement de variable de la forme \_\_VAR\_\_.

La génération de l'image se fait par un script Shell qui monte l'image ISO, crée un overlay pour ne pas avoir à recopier le contenu de l'ISO, crée les fichiers spécifiques et copie les scripts Preseed et tous les fichiers à installer sur la machine.

```
mount -o loop "${ISOFILE}" "${ISOMOUNTDIR}"
mount -t overlay -o lowerdir=${ISOMOUNTDIR},upperdir=${UPPERDIR},
                 workdir=${WORKINGDIR} overlay ${NEWISODIR}
...
mkisofs -r -V "${HOST} Ubuntu" \
        -cache-inodes \
        -J -l -b isolinux/isolinux.bin \
        -c isolinux/boot.cat -no-emul-boot \
        -boot-load-size 4 -boot-info-table \
        -o $ISOBUILD $NEWISODIR
```

Cette image est copiée sur un partage NFS qui permet de la mettre à disposition des iDracs.

### 2.3.3 Déploiement des images

```
...

racadm -r ${DRACHOST} -u root -p ${PASS} remoteimage -c -u user
-p pass -l "NFS\_IMG\_PATH"

racadm -r ${DRACHOST} -u root -p ${PASS} set
idrac.serverboot.BootOnce 1

racadm -r ${DRACHOST} -u root -p ${PASS} set
idrac.serverboot.FirstBootDevice VCD-DVD

racadm -r ${DRACHOST} -u root -p ${PASS} remoteimage -s

racadm -r ${DRACHOST} -u root -p ${PASS} serveraction powercycle
```

Cette étape passée, la machine est installée et utilisable. Le processus prend environ 15 minutes et peut être lancé en parallèle sur un ensemble de nœuds. Il va maintenant falloir déployer OpenStack. Kolla va rendre le processus rapide, efficace et reproductible.

## 2.4 Kolla

### 2.4.1 Choix de Kolla

Il existe de multiples manières de déployer OpenStack. Parmi elles, nous allons aborder TripleO, OpenStack Ansible et Kolla. Depuis, d'autres solutions sont apparues, que nous n'aborderons pas ici.

TripleO est un projet très intéressant qui part du principe que si OpenStack est une bonne solution pour gérer notre infrastructure, on peut l'utiliser pour déployer… OpenStack ! On déploie donc le cluster OpenStack dans OpenStack (d'où le nom de TripleO : OpenStack On OpenStack), à l'aide de Heat, l'orchestrateur OpenStack et de scripts Puppet. Cette solution a été choisie par exemple par RedHat pour RDO, la solution de déploiement d'OpenStack. Si cette solution peut sembler séduisante, la configuration manque de flexibilité, le système d'extensions limitant les possibilités de modification. De plus lors d'une modification de configuration, on n'a pas de contrôle fin sur les nœuds qui doivent recevoir la nouvelle configuration. Tous les nœuds sont reconfigurés, ce qui est souvent long et inutile. Enfin, la gestion de l'*undercloud*, apporte au final une couche de complexité supplémentaire.

OpenStack Ansible est une solution plus simple, se basant simplement sur des scripts Ansible. C'est une solution stable et robuste, mais qui dépend fortement des systèmes hôtes.

Le projet Kolla fournit des images de conteneurs Docker et les outils de déploiement qui permettent de déployer OpenStack de manière simple et rapide. Il permet un passage à l'échelle facile et est agnostique quant aux systèmes des hôtes, qui doivent simplement fournir SSH et le démon Docker. L'utilisation de conteneurs permet des déploiements reproductibles. A l'époque de notre choix deux solutions exploitant ces conteneurs étaient disponibles : Kolla Ansible, qui se base sur des scripts Ansible, et Kolla Kubernetes, qui déploie les conteneurs sur un cluster Kubernetes. Cette dernière solution nous a paru trop complexe, et nous risquions d'accumuler une dette technique inutile. Nous avons donc choisi la solution Ansible, ce qui s'est avéré être un bon choix, puisque Kolla Kubernetes a été depuis abandonné au profit de Kubernetes Helm.

Le projet Kolla permet aussi de déployer une infrastructure Ceph. Mais nous avons choisi de séparer OpenStack et Ceph pour une meilleure flexibilité, ce dernier pouvant être utilisé par d'autres projets pour du stockage objet ou système de fichiers.

Kolla permet de configurer l'accès aux stockages Ceph et SolidFire pour Openstack.

Passée une phase de prise en main, l'utilisation de Kolla apporte de nombreux bénéfices : la reproductibilité permet d'avoir des infrastructures de tests et de production identiques ; l'ensemble de la configuration peut être stockée dans un dépôt de source et nous pouvons bénéficier des méthodes qui vont avec. Enfin, les montées en version, réputées difficiles, sont directement gérées par Kolla.

### 2.4.2 En pratique

Pour configurer Kolla, trois fichiers suffisent : un fichier d'inventaire `inventory.yml`.

```
[control]
openstack-[01:03].ex.fr    ansible_user=root
ansible_ssh_common_args='-o StrictHostKeyChecking=no'

[network]
openstack-[01:03].ex.fr    ansible_user=root
ansible_ssh_common_args='-o StrictHostKeyChecking=no'

[external-compute]
openstack-c-[01:03].ex.fr  ansible_user=root
ansible_ssh_common_args='-o StrictHostKeyChecking=no'

[storage]
openstack-[01:03].ex.fr    ansible_user=root
```

Le fichier `globals.yml` qui permet de configurer le réseau, les services à installer, la haute disponibilité, *etc*.

```
kolla_base_distro: "centos"
kolla_install_type: "binary"
openstack_release: "queens"

kolla_internal_vip_address: "x.x.x.x"
kolla_external_vip_address: "y.y.y.y"
kolla_external_fqdn: "cloud.ex.fr"

network_interface: "eno1"
tunnel_interface: "enp59s0f1"
neutron_external_interface: "eno2"

enable_ceph: "no"
enable_cinder: "yes"
enable_cinder_backup: "no"
enable_cinder_backend_iscsi: "yes"
enable_neutron_dvr: "yes"
```

Le fichier `password.yml` qui contient les mots de passe utilisés pour la communication entre les services.

```
kolla-genpwd -p passwords.yml
```

La dernière étape est de lancer l'installation.

```
kolla-ansible -i inventory.yml --configdir . --passwords
passwords.yml bootstrap-servers

kolla-ansible -i inventory.yml --configdir . --passwords
passwords.yml prechecks

kolla-ansible -i inventory.yml --configdir . --passwords
passwords.yml pull

kolla-ansible -i inventory.yml --configdir . --passwords
passwords.yml  -e certdir=$certdir deploy
```

Si les variables Kolla, ont été correctement déclarées, le cluster OpenStack est fonctionnel.

## 2.5 Post installation

Des scripts Shell réalisent la post-installation. Ainsi, la création des saveurs (*flavor*) des futures machines virtuelles est automatisée en se basant sur des cas standards : 1 cœur + X RAM, 2 cœurs… Il en est de même pour la création des images de quelques OS prédéfinis (Debian, CentOS, *etc.*). Le futur utilisateur peut ainsi facilement choisir les ressources dont il a besoin ainsi que son OS, avec des systèmes préalablement validés.

Un script permet la création aisée de nouveaux projets, en particulier en affectant les plages d'adresses IP privées et publiques, et en configurant automatiquement la solution de SDN à laquelle l'infrastructure est raccordée.

L'ensemble des fichiers est historisé par la forge GitLab gérée par un autre comité technique de l'UMS GRICAD. Une machine virtuelle Debian GNU/Linux minimale a été déployée sur le service de virtualisation VMware du Campus (WINTER [2]) afin d'avoir un point central pour lancer l'installation complète. En cas de simple reconfiguration du nuage, la solution Kolla optimise les containers à mettre à jour, et les services à relancer. Une réinstallation complète du cluster depuis ce nœud en partant de zéro prend environ 30 minutes. L'intégration d'un nouveau nœud de calcul prend un temps équivalent.

## 2.6 Mise à jour de la plateforme

### 2.6.1 Kolla / OpenStack

C'est la partie la plus sensible. Malgré le déploiement de la HA sur l'ensemble des services par Kolla, la mise à jour des contrôleurs occasionne des pertes de connectivité réseau pour les VMs. Ces déconnexions peuvent atteindre quelques minutes.

Lors de la mise à jour des *computes*, nous n'observons pas de problème, car nous passons les machines en maintenance. Toutes les VMs sont migrées à chaud sans interruption de service, puis la machine mise à jour. Une fois la mise à jour terminée, les VMs sont alors redistribuées sur les nœuds pour équilibrer la charge.

Le processus de mise en maintenance d'un *compute* passe par un script que nous avons développé :

```
openstack compute service set --disable \
                              --disable-reason "update" \
                              "${host}" \
                              nova-compute

# For each state: ACTIVE, SHUTOFF, SUSPENDED
for key in ${!srv[@]}; do
   openstack server migrate --live \{} \
     "${active_computes[$cc]}" "${srv[$key]}"
done
```

#### 2.6.2 Ceph

La triple réplication des données et les trois contrôleurs permettent de réaliser les mises à jour à chaud sans interruption de service. Le redémarrage des contrôleurs est transparent, car l'un des deux autres prend le relais automatiquement. Pour les serveurs de stockage, il suffit d'exécuter les lignes suivantes pour permettre de relancer une des machines sans casser le cluster :

```
# OSD that do not respond are put out of the cluster
ceph osd set noout

# We do not rebalance the cluster
ceph osd set norebalance

# After restarting the storage server,
# you put everything back in order
ceph osd unset noout
ceph osd unset norebalance
```

Depuis le lancement de la plateforme Nova, le cluster Ceph est passé en Nautilus sous Ubuntu 18.04 LTS. Ainsi nous n'avons pas de changement de version d'OS ou de Ceph à prévoir avant le renouvellement du matériel.

## 3 Cas d'usage

Au moment où nous écrivons ces lignes, la plateforme héberge 18 projets, pour un total de 128 cores utilisés, 400Go de RAM, et une dizaine de To de stockage. Les projets arrivent petit à petit et abordent des sujets très divers :

- Système Informatique Distribué d'Évaluation en Santé (SIDES) ;
- intégration continue en liaison avec la plateforme GitLab ;
- plateforme de test pour l'ordonnanceur de containers Kubernetes ;
- logiciels de calcul et de visualisation (mutualisation de licences logicielles) ;
- plateforme de test et d'évaluation d'infrastructure Beegfs ;
- Notebook Jupyter au travers de la mise à disposition de serveurs Jupyterhub (portail web). Un utilisateur peut choisir le service qu'il désire (examens en ligne, challenges, TP, formation…) et le langage (Python, R, Julia…). En fonction du nombre de participants demandé, une machine virtuelle optimisée est automatiquement allouée et configurée. La machine libère ses ressources au dernier utilisateur sortant et se relance si besoin.

## 4  Perspectives

Maintenant que nous avons une plateforme opérationnelle, nous voulons permettre à tout le monde de s'en servir, informaticiens et non-informaticiens. Pour cela nous avons commencé à regarder comment fournir des services clef en main pour la collecte, l'analyse, le calcul et la mise à disposition de données, via des plateformes basées par exemple sur Elasticsearch, Logstash, Kibana.

Il est aussi nécessaire d'étudier les solutions de chiffrement des infrastructures de stockage dans le cadre du RGPD et de la PPST, pour des projets qui hébergent des données sensibles et/ou à caractère personnel.

Un autre axe de recherche est la mise à disposition de nœuds GPU pour le calcul. Nous sommes en train d'étudier les architectures matérielles les plus adaptées pour l'intégration avec OpenStack.

Il nous faut aussi consolider l'utilisation de Nova pour les enseignements, qui sont de plus en plus demandeurs de ces technologies qui font désormais partie des cursus.

L'usage actuel de Nova fait ressortir un besoin de mutualisation des licences logicielles par les laboratoires pour le calcul et la recherche. Ces licences étant très chères et les installations parfois compliquées, cela permet d'optimiser les temps d'utilisation.

Nous désirons étudier comment permettre l'hébergement de services non critiques, comme des serveurs web de laboratoires, sans pour autant pénaliser les besoins en traitement de données sur la plateforme.

Actuellement, toute notre infrastructure se situe dans une seule salle du data-centre Imag. Nous souhaitons mettre en place du multi-site pour augmenter la résilience de Nova.

Nous travaillons aussi en collaboration avec France Grille pour à terme intégrer Nova dans le réseau national de cloud.

Enfin, pour compléter la documentation, nous devons monter des formations sur les possibilités et l'utilisation de Nova.

Concernant l'exploitation de Nova, nous devons améliorer l'automatisation des procédures de validation de mises à jour, ainsi que de leur déploiement.

## 5  Conclusion

Le nuage OpenStack est aujourd'hui opérationnel. Le savoir-faire a été acquis, capitalisé et est documenté. L'ensemble du nuage peut être reconfiguré/déployé rapidement. Construit sur des composants matériels robustes et hétérogènes, ce nuage peut donc être facilement étendu avec de nouveaux nœuds.

Le modèle fédéral porté par l'UMS GRICAD a permis la réalisation de ce projet d'ampleur. Au-delà des aspects technologiques, la collaboration autour d'un projet très concret avec des personnes dispersées sur le Campus est une expérience professionnelle et personnelle enrichissante.

# Bibliographie